\documentclass[runningheads]{llncs}
\usepackage[T1]{fontenc}
\usepackage{graphicx,verbatim}

\usepackage{booktabs}
\usepackage{amsmath}
\usepackage{amssymb}
\usepackage{amsfonts}
\usepackage{multirow}
\usepackage[table]{xcolor}

\usepackage{bbding}

\definecolor{lightgray}{gray}{0.92}
\definecolor{first}{RGB}{191, 225, 201} 
\definecolor{second}{RGB}{227, 237, 185} 
\definecolor{third}{RGB}{254, 250, 194} 
\definecolor{stab}{RGB}{232, 241, 252} 
\newcommand{\err}[1]{{\tiny $\pm$#1}}

\usepackage{hyperref}

\begin{document}
\title{Unsupervised Causal Prototypical Networks for De-biased Interpretable Dermoscopy Diagnosis}
%

\author{Junhao Jia\inst{1,2} \and
Yueyi Wu\inst{2} \and
Huangwei Chen\inst{1,2} \and
Haodong Jing\inst{3} \and \\
Haishuai Wang\inst{1} \and
Jiajun Bu\inst{1}\textsuperscript{\smash{(\Envelope)}} \and
Lei Wu\inst{1}\textsuperscript{\smash{(\Envelope)}}
}
\authorrunning{F. Author et al.}
%
\institute{
Zhejiang Key Laboratory of Accessible Perception and Intelligent Systems, College
of Computer Science and Technology, Zhejiang University
\and Hangzhou Dianzi University \quad \textsuperscript{3} Xi'an Jiaotong University \\
\email{\{bjj, shenhai1895\}@zju.edu.cn}}
  
\maketitle              
\begin{abstract}
Despite the success of deep learning in dermoscopy image analysis, its inherent black-box nature hinders clinical trust, motivating the use of prototypical networks for case-based visual transparency. However, inevitable selection bias in clinical data often drives these models toward shortcut learning, where environmental confounders are erroneously encoded as predictive prototypes, generating spurious visual evidence that misleads medical decision-making. To mitigate these confounding effects, we propose CausalProto, an Unsupervised Causal Prototypical Network that fundamentally purifies the visual evidence chain. Framed within a Structural Causal Model, we employ an Information Bottleneck-constrained encoder to enforce strict unsupervised orthogonal disentanglement between pathological features and environmental confounders. By mapping these decoupled representations into independent prototypical spaces, we leverage the learned spurious dictionary to perform backdoor adjustment via do-calculus, transforming complex causal interventions into efficient expectation pooling to marginalize environmental noise. Extensive experiments on multiple dermoscopy datasets demonstrate that CausalProto achieves superior diagnostic performance and consistently outperforms standard black box models, while simultaneously providing transparent and high purity visual interpretability without suffering from the traditional accuracy compromise.


\end{abstract}
\section{Introduction}

While deep learning has achieved remarkable success in the automated analysis of dermoscopy images, its reliance on high-dimensional feature representations and massive parameter scales renders these models inherently opaque~\cite{esteva2017dermatologist}. This black-box nature severely restricts their deployment in safety-critical clinical environments~\cite{rudin2019stop}, where practitioners require both high diagnostic accuracy and transparent decision-making evidence that aligns with expert comprehension. To bridge this gap, prototypical learning has emerged as a promising solution in medical vision tasks~\cite{chen2019looks,barnett2021case}. By adopting a case-based reasoning paradigm, this architecture transforms abstract feature mappings into intuitive similarity metrics between input samples and learned pathological prototypes, thereby inherently mirroring human cognitive logic as revealed by psychological studies~\cite{nosofsky1986attention}.

\begin{figure}
    \centering
    \includegraphics[width=1\linewidth]{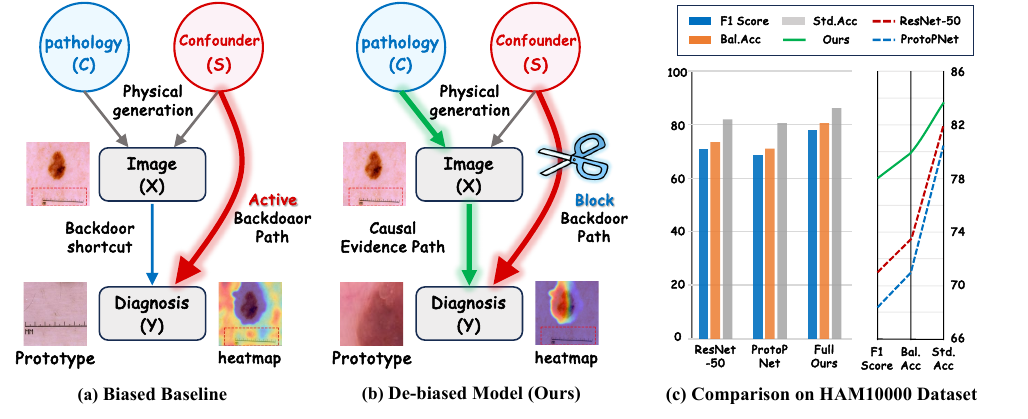}
    \caption{The underlying Structural Causal Model and performance overview.}
    \label{fig:intro}
\end{figure}

However, this interpretability mechanism exhibits critical vulnerabilities when confronted with the pervasive selection bias inherent to real-world clinical data~\cite{geirhos2020shortcut}, undermining its potential to foster clinical trust. As illustrated in Fig.~\ref{fig:intro}, traditional prototypical networks are highly susceptible to shortcut learning, frequently encoding ubiquitous dermoscopic artefacts into biased prototypes rather than capturing the genuine pathological essence of a lesion~\cite{winkler2019association,winkler2021association}. This reliance on environmental confounders establishes an active backdoor path, where spurious visual evidence dictates the decision-making process and renders the diagnosis fundamentally unreliable~\cite{castro2020causality}. From the perspective of a Structural Causal Model~\cite{neuberg2003causality}, this diagnostic failure is rooted in confounding effects during the observational data generation process. By merely fitting observational conditional probabilities, traditional Empirical Risk Minimization frameworks inherently exploit the spurious correlations perpetuated by this backdoor path~\cite{arjovsky2019invariant,scholkopf2021toward}.

To rectify these diagnostic failures, we introduce CausalProto, an unsupervised causal prototypical network designed to decouple pathological features from environmental confounders via an information bottleneck~\cite{tishby2000information,alemi2017deep}, while employing do-calculus for backdoor adjustment~\cite{neuberg2003causality} to marginalize spurious noise. Our contributions are summarized as follows:
\underline{First,} we rigorously define the spurious evidence generation mechanism in medical vision tasks, revealing the critical vulnerability of prototypical networks to confounding factors~\cite{geirhos2020shortcut}.
\underline{Second,} we achieve strict orthogonal disentanglement between pathological and environmental features via variational mutual information upper bound approximation, bypassing the need for environmental annotations~\cite{cheng2020club}.
\underline{Third,} we introduce an unsupervised confounding prototype library as a causal intervention dictionary and employ do-calculus for efficient expectation pooling, thereby marginalizing spurious noise and restoring reliable causal reasoning.
\underline{Fourth,} extensive experiments on multiple dermoscopy datasets demonstrate that CausalProto provides transparent visual interpretability while simultaneously achieving superior diagnostic accuracy, overcoming the traditional accuracy-interpretability trade-off.

\begin{figure}[htbp]
    \centering
    \includegraphics[width=0.99\linewidth]{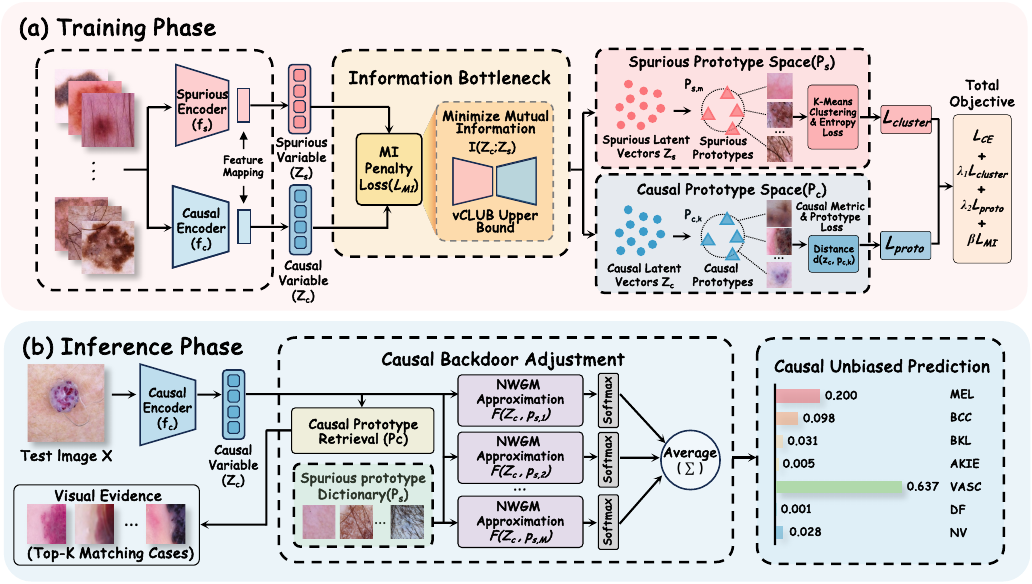}
    \caption{The framework of CausalProto. (a) Training Phase: A dual-branch network disentangles causal ($Z_C$) and spurious ($Z_S$) features via mutual information minimization. Distinct causal ($P_C$) and spurious ($P_S$) prototype spaces are optimized for interpretable representation learning. (b) Inference Phase: Causal backdoor adjustment is performed by marginalizing over the learned spurious prototype dictionary ($P_S$) via NWGM pooling to achieve de-biased interventional prediction $P(Y|do(X))$.}
    \label{fig:pipeline}
\end{figure}

\section{Methodology}

\subsection{SCM Formulation and Framework Architecture}

Given an observed image $X \in \mathcal{X}$ annotated with a diagnostic label $Y \in \mathcal{Y}$, our objective is to develop an unsupervised causal prototypical network (CausalProto) capable of performing de-biased interpretable dermoscopy diagnosis. To formally address the issue of shortcut learning, we postulate that the data generation mechanism follows a Structural Causal Model (SCM). We assume $X$ is generated by two underlying, mutually independent latent factors: causal factors $C \in \mathcal{C}$ representing genuine pathological attributes and spurious factors $S \in \mathcal{S}$ representing environmental artifacts. In real-world clinical datasets, selection biases establish a confounding backdoor path $Y \leftarrow S \rightarrow X$, which traditional Empirical Risk Minimization (ERM) models greedily exploit by merely modeling the observational conditional probability $P(Y|X)$~\cite{arjovsky2019invariant,scholkopf2021toward}.

To sever these spurious correlations and compel the model to learn the true interventional probability $P(Y|do(X))$, the CausalProto framework is designed with three core mapping processes, as illustrated in Fig.~\ref{fig:pipeline}. The first functions $f_c(\cdot)$ and $f_s(\cdot)$, referred to as the dual-branch encoders, learn to map the input image $X$ into disentangled causal latent variables $Z_C$ and spurious latent variables $Z_S$. The second process maps these representations into two discrete prototypical spaces: a causal prototype library $P_C$ to capture genuine pathological patterns, and a spurious prototype library $P_S$ to model environmental artifacts. Finally, the inference function performs causal backdoor adjustment, predicting the label $Y$ by utilizing Normalized Weighted Geometric Mean (NWGM) pooling to marginalize over the learned spurious dictionary $P_S$.

\subsection{Unsupervised Disentanglement via Information Bottleneck}

Given the input image $X$, we deploy parallel dual-branch encoders $f_c(\cdot)$ and $f_s(\cdot)$ to extract the causal latent representation $Z_C = f_c(X)$ and the spurious latent representation $Z_S = f_s(X)$, respectively. To guarantee strict orthogonality and mutual independence between these two feature spaces without relying on environmental annotations, we enforce an Information Bottleneck (IB) constraint. The objective is to forcefully minimize the mutual information (MI) between $Z_C$ and $Z_S$, formally defined as:
\begin{equation}
I(Z_C ; Z_S) = \iint p(z_c, z_s) \log \frac{p(z_c, z_s)}{p(z_c)p(z_s)} dz_c dz_s
\end{equation}

Since computing exact MI in a high-dimensional continuous space is computationally intractable, we adopt the Variational Contrastive Log-Ratio Upper Bound (vCLUB) approximation~\cite{cheng2020club}. Let $q_\theta(z_c|z_s)$ denote a variational approximation network parameterized by a Multi-Layer Perceptron (MLP). The upper bound is formulated as:
\begin{equation}
I(Z_C; Z_S) \le \mathbb{E}_{p(z_c, z_s)} \left[ \log q_\theta(z_c|z_s) \right] - \mathbb{E}_{p(z_c)p(z_s)} \left[ \log q_\theta(z_c|z_s) \right]
\end{equation}

During training, we employ an alternating optimization strategy. First, we maximize $q_\theta$ to approximate the true conditional distribution $p(z_c|z_s)$. Subsequently, fixing $q_\theta$, we optimize the encoders $f_c$ and $f_s$ to minimize this bound. The corresponding empirical MI penalty loss over a batch of $N$ samples is defined as:
\begin{equation}
\mathcal{L}_{MI} = \frac{1}{N} \sum_{i=1}^N \left[ \log q_\theta(z_{c,i}|z_{s,i}) - \frac{1}{N} \sum_{j=1}^N \log q_\theta(z_{c,j}|z_{s,i}) \right]
\end{equation}

\subsection{Causal Prototypical Metric for Interpretable Reasoning}

To achieve intrinsic interpretability via case-based reasoning, we establish two independent latent prototype spaces: a causal library $P_C = \{p_{c,k}\}_{k=1}^K$ capturing genuine pathological patterns, and a spurious library $P_S = \{p_{s,m}\}_{m=1}^M$ modeling environmental artifacts.

For a derived causal feature $Z_C$, the diagnostic probability is determined by its proximity to the valid causal prototypes within each class. Using the Euclidean distance $d(\cdot, \cdot)$, the predictive distribution is formulated as:
\begin{equation}
P(Y | Z_C) = \frac{\sum_{k \in \mathcal{K}_y} \exp(-d(Z_C, p_{c,k}))}{\sum_{y'} \sum_{k' \in \mathcal{K}_{y'}} \exp(-d(Z_C, p_{c,k'}))}
\end{equation}

To guarantee that the classification basis $P_C$ represents verifiable clinical examples rather than abstract vectors, we enforce a strict prototype projection constraint. Specifically, each causal prototype $p_{c,k}$ is periodically mapped to the nearest latent representation of a real training image instance from its corresponding class~\cite{chen2019looks,jia2025geodesic}:
\begin{equation}
p_{c,k} \leftarrow \underset{z_c \in \{f_c(x) | (x,y) \in \mathcal{D}_{train}\}}{\arg\min} d(z_c, p_{c,k})
\end{equation}

\subsection{Backdoor Adjustment via do-calculus}

During inference, standard conditional predictions $P(Y|X)$ remain susceptible to spurious correlations propagated through the backdoor path $S \rightarrow X \rightarrow Y$. To isolate the true causal effect, we block this path using Pearl’s do-calculus. Utilizing the disentangled causal representation $Z_C$, the interventional distribution is formulated via stratified adjustment over the confounding space $\mathcal{S}$:
\begin{equation}
P(Y | do(X)) = \int_{\mathcal{S}} P(Y | Z_C, z_s) P(z_s) dz_s
\end{equation}

Since direct integration over the continuous, high-dimensional space $\mathcal{S}$ is intractable, we leverage the unsupervisedly learned spurious dictionary $P_S = \{p_{s,m}\}_{m=1}^M$, which inherently spans the manifold of environmental confounders. This enables us to approximate the continuous causal adjustment as a tractable expectation pooling over this discrete dictionary via the Normalized Weighted Geometric Mean (NWGM)~\cite{xu2015show,wang2020visual}:
\begin{equation}
P(Y | do(X)) \propto \mathbb{E}_{p_s \sim P_S} \left[ P(Y | Z_C, p_s) \right] \approx \frac{1}{M} \sum_{m=1}^M \text{Softmax} \left( \mathcal{F}(Z_C, p_{s,m}) \right)
\end{equation}

where $\mathcal{F}$ denotes the feature fusion network. This mechanism mathematically marginalizes specific environmental artifacts by averaging the Softmax probability distributions across the diverse contexts captured in $P_S$, ensuring the final diagnosis is robust, de-biased, and strictly driven by the genuine pathological evidence in $P_C$.

\subsection{Overall Objective}

To optimize the CausalProto framework end-to-end, the total objective function is designed to jointly balance predictive accuracy, prototypical interpretability, and feature disentanglement:
\begin{equation}
\mathcal{L} = \mathcal{L}_{CE}(Y, \hat{Y}_{do}) + \lambda_1 \mathcal{L}_{cluster}(Z_S, P_S) + \lambda_2 \mathcal{L}_{proto}(Z_C, P_C) + \beta \mathcal{L}_{MI}(Z_C, Z_S)
\end{equation}

where $\mathcal{L}_{CE}$ denotes the cross-entropy loss for the causal intervention prediction. The term $\mathcal{L}_{cluster}$ enforces the semantic diversity of the spurious prototype dictionary, while $\mathcal{L}_{proto}$ regularizes the latent space to align input features with corresponding causal class prototypes. The mutual information penalty $\mathcal{L}_{MI}$ guarantees orthogonal feature disentanglement, with $\lambda_1$, $\lambda_2$, and $\beta$ serving as balancing hyperparameters.

\section{Experiments}

\subsection{Experimental Setup}

\subsubsection{Datasets and metrics:} To comprehensively evaluate the diagnostic performance and clinical applicability of CausalProto across diverse real-world settings, we utilize three public skin lesion datasets: HAM10000~\cite{tschandl2018ham10000} (10,015 dermoscopic images across 7 classes with severe imbalance), ISIC 2019~\cite{codella2019skin} (approximately 25k images, 8 classes), and PAD-UFES-20~\cite{pacheco2020pad} (a clinical smartphone dataset with 2,298 images from 1,373 patients). Performance is evaluated using Accuracy (Acc), Balanced Accuracy (BAcc), and Macro-F1 (F1). Results are reported as mean $\pm$ standard deviation across five random seeds.

\subsubsection{Implementation details:} All models are implemented using PyTorch and trained on a single NVIDIA RTX 3090 GPU. Images are resized to 224 $\times$ 224. We employ an ImageNet-pretrained ResNet-50 as the default backbone. Optimization is performed using Adam with an initial learning rate of $1 \times 10^{-4}$ (cosine decay) and a weight decay of $1 \times 10^{-4}$. Models are trained for $100$ epochs with a batch size of $32$. Optimal hyperparameters ($\lambda_1 = 0.1$, $\lambda_2 = 0.1$, $\beta = 0.5$) and dictionary sizes ($K=10$, $M=50$) are selected via grid search on the validation set. We adopt official data splits and report results across five random seeds. Statistical significance is assessed via paired t-tests ($p < 0.05$).

\subsubsection{Baselines:} We compare CausalProto against three families of methods: (1) Black-box ERM models, including ResNet-50~\cite{he2016deep}, EfficientNet~\cite{tan2019efficientnet}, and Swin-T~\cite{liu2021swin}, to establish standard predictive baselines. (2) Prototype-based interpretable models, comprising ProtoPNet~\cite{chen2019looks}, LGProto~\cite{santiago2023global}, PIP-Net~\cite{nauta2023pip} and Proto-RSet~\cite{donnelly2025rashomon}, which provide part-based or global visual explanations but remain vulnerable to confounding. (3) Robust Causal Representation models, including Group DRO~\cite{sagawadistributionally}, FactorVAE~\cite{kim2018disentangling}, and CausalVAE~\cite{yang2021causalvae}, which aim to mitigate spurious correlations via robust optimization or factorized disentanglement. All baseline methods are strictly tuned following their original hyperparameter configurations to ensure a fair comparison.

\begin{table}[h]
\centering
\caption{Quantitative comparison of diagnostic performance and key ablations.}
\label{tab1}

\setlength{\tabcolsep}{1.6pt}
\resizebox{\textwidth}{!}{
\begin{tabular}{l|ccc|ccc|ccc}
\toprule
\multirow{2}{*}{\textbf{Method}} 
& \multicolumn{3}{c|}{\textbf{HAM10000}} 
& \multicolumn{3}{c|}{\textbf{ISIC 2019}} 
& \multicolumn{3}{c}{\textbf{PAD-UFES-20}} \\
\cmidrule(lr){2-4} \cmidrule(lr){5-7} \cmidrule(lr){8-10}
& \textit{Acc} & \textit{BAcc} & \textit{F1}
& \textit{Acc} & \textit{BAcc} & \textit{F1}
& \textit{Acc} & \textit{BAcc} & \textit{F1} \\
\midrule

\multicolumn{10}{l}{\textbf{Black-box ERM Models}} \\
ResNet-50     & 82.0\err{1.2} & 73.5\err{1.6} & 71.0\err{1.9} & 78.0\err{1.4} & 68.0\err{1.7} & 65.0\err{2.0} & 72.0\err{2.1} & 60.0\err{2.7} & 57.0\err{2.9} \\
EfficientNet   & 83.5\err{1.1} & 75.0\err{1.5} & 72.5\err{1.8} & 79.5\err{1.3} & 70.0\err{1.6} & 67.0\err{1.9} & 73.0\err{2.0} & 61.5\err{2.6} & 58.5\err{2.7} \\
Swin-T        & 83.0\err{1.1} & 74.2\err{1.6} & 72.0\err{1.8} & 80.0\err{1.2} & 70.8\err{1.6} & 67.8\err{1.8} & 72.5\err{2.3} & 60.8\err{2.9} & 57.8\err{3.1} \\
\midrule

\multicolumn{10}{l}{\textbf{Prototype-baesd Interpretable Models}} \\
ProtoPNet      & 80.5\err{1.6} & 71.0\err{2.0} & 68.5\err{2.3} & 76.5\err{1.6} & 66.0\err{2.0} & 63.0\err{2.3} & 70.0\err{2.6} & 58.0\err{3.2} & 54.5\err{3.4} \\
LGProto        & 82.8\err{1.4} & 74.0\err{1.8} & 71.5\err{2.0} & 79.0\err{1.4} & 69.8\err{1.8} & 66.8\err{2.0} & 74.5\err{2.0} & 63.5\err{2.5} & 60.5\err{2.7} \\
PIP-Net        & 83.2\err{1.4} & 74.5\err{1.8} & 72.0\err{2.0} & 80.2\err{1.2} & 71.2\err{1.6} & 68.5\err{1.8} & 72.0\err{2.6} & 60.0\err{3.3} & 56.5\err{3.6} \\
Proto-RSet     & 82.5\err{1.5} & 73.5\err{1.9} & 71.0\err{2.1} & 79.4\err{1.3} & 70.5\err{1.7} & 67.5\err{1.9} & 72.5\err{2.4} & 61.0\err{3.0} & 58.0\err{3.1} \\
\midrule

\multicolumn{10}{l}{\textbf{Robust Causal Representation Models}} \\
Group DRO     & 82.2\err{1.2} & 76.0\err{1.4} & 73.5\err{1.6} & 78.8\err{1.4} & 71.2\err{1.5} & 68.3\err{1.6} & 72.2\err{2.1} & 63.0\err{2.2} & 60.0\err{2.3} \\
FactorVAE      & 83.2\err{1.1} & 75.8\err{1.5} & 73.5\err{1.6} & 79.4\err{1.3} & 70.6\err{1.5} & 67.8\err{1.7} & 73.4\err{2.0} & 62.5\err{2.3} & 59.6\err{2.4} \\
CausalVAE     & 83.5\err{1.1} & 76.4\err{1.4} & 74.1\err{1.6} & 79.8\err{1.2} & 71.2\err{1.5} & 68.4\err{1.6} & 73.8\err{1.9} & 63.0\err{2.2} & 60.2\err{2.3} \\
\midrule

\multicolumn{10}{l}{\textbf{Key Ablations of our Model}} \\
w/o MI                    & 85.0\err{0.9} & 75.2\err{1.3} & 72.5\err{1.6} & 81.0\err{1.0} & 69.0\err{1.5} & 65.5\err{1.7} & 77.0\err{1.7} & 64.0\err{2.1} & 60.8\err{2.2} \\
w/o Cluster                   & 85.6\err{0.9} & 78.4\err{1.2} & 76.0\err{1.4} & 81.7\err{0.9} & 72.8\err{1.3} & 70.0\err{1.5} & 77.9\err{1.5} & 67.5\err{1.9} & 65.0\err{2.0} \\
w/o do-calc                    & 84.8\err{1.0} & 77.0\err{1.3} & 74.2\err{1.6} & 81.2\err{1.0} & 71.0\err{1.5} & 68.0\err{1.7} & 77.2\err{1.7} & 66.0\err{2.1} & 63.0\err{2.2} \\
Shared Proto                    & 84.0\err{1.0} & 75.8\err{1.4} & 73.0\err{1.7} & 80.5\err{1.0} & 69.5\err{1.6} & 66.2\err{1.8} & 76.0\err{1.8} & 63.0\err{2.3} & 59.5\err{2.4} \\
\midrule

\rowcolor{lightgray}
\textbf{Full (Ours)}      & \textbf{86.2}\err{0.8} & \textbf{80.5}\err{1.0} & \textbf{78.0}\err{1.2}
                          & \textbf{82.0}\err{0.9} & \textbf{75.0}\err{1.1} & \textbf{72.0}\err{1.3}
                          & \textbf{78.5}\err{1.4} & \textbf{70.0}\err{1.8} & \textbf{67.5}\err{1.9} \\
\bottomrule
\end{tabular}
}
\end{table}

\begin{table}[h]
\centering
\fontsize{8pt}{10pt}\selectfont 
\caption{Prototype purity and disentanglement quality. NMI is the estimated $I(Z_C;Z_S)$ upper bound ($\downarrow$). Purity measures the class-consistency of causal prototypes ($\uparrow$). Div is the spurious assignment entropy indicating confounder coverage ($\uparrow$).}
\label{tab2}

\setlength{\tabcolsep}{2.5pt} 

\begin{tabular}{l|ccc|ccc|ccc}
\toprule
\multirow{2}{*}{\textbf{Method}}
& \multicolumn{3}{c|}{\textbf{HAM10000}}
& \multicolumn{3}{c|}{\textbf{ISIC 2019}}
& \multicolumn{3}{c}{\textbf{PAD-UFES-20}} \\
\cmidrule(lr){2-4} \cmidrule(lr){5-7} \cmidrule(lr){8-10}
& \textit{NMI$\downarrow$} & \textit{Purity$\uparrow$} & \textit{Div$\uparrow$}
& \textit{NMI$\downarrow$} & \textit{Purity$\uparrow$} & \textit{Div$\uparrow$}
& \textit{NMI$\downarrow$} & \textit{Purity$\uparrow$} & \textit{Div$\uparrow$}\\
\midrule

\multicolumn{10}{l}{\textbf{Baselines}} \\
ProtoPNet            & -             & $0.52 \pm 0.03$          & -             & -             & $0.58 \pm 0.02$          & -             & -             & $0.49 \pm 0.04$          & - \\
PIP-Net              & -             & $0.58 \pm 0.02$          & -             & -             & $0.62 \pm 0.02$          & -             & -             & $0.54 \pm 0.03$          & - \\
\midrule

\multicolumn{10}{l}{\textbf{Key ablations}} \\
w/o MI                & 0.35          & $0.55 \pm 0.03$          & 2.00          & 0.31          & $0.60 \pm 0.03$          & 2.15          & 0.38          & $0.52 \pm 0.04$          & 2.25 \\
w/o Cluster          & 0.10          & $0.77 \pm 0.02$          & 1.50          & 0.09          & $0.80 \pm 0.01$          & 1.65          & 0.12          & $0.74 \pm 0.02$          & 1.80 \\
w/o do-calc          & 0.07          & $0.81 \pm 0.01$          & 3.18          & 0.06          & $0.84 \pm 0.01$          & 3.28          & 0.08          & $0.79 \pm 0.02$          & 3.58 \\
Shared Proto         & 0.23          & $0.62 \pm 0.03$          & 2.18          & 0.20          & $0.66 \pm 0.02$          & 2.35          & 0.28          & $0.58 \pm 0.03$          & 2.50 \\
\midrule

\rowcolor{lightgray}
\textbf{Full (Ours)} & \textbf{0.07} & \textbf{0.82$\pm$0.01} & \textbf{3.20} & \textbf{0.06} & \textbf{0.85$\pm$0.01} & \textbf{3.30} & \textbf{0.08} & \textbf{0.80$\pm$0.01} & \textbf{3.60} \\
\bottomrule
\end{tabular}
\end{table}

\begin{figure}
    \centering
    \includegraphics[width=\linewidth]{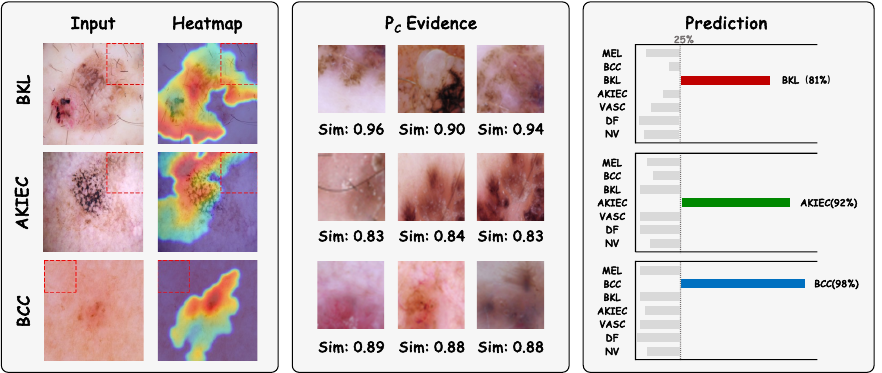}
    \caption{Qualitative visualization of CausalProto's de-confounded reasoning.}
    \label{fig:exp}
\end{figure}

\subsection{Results and Analysis}

\subsubsection{Comparative analysis:} As shown in Table~\ref{tab1}, CausalProto consistently achieves state-of-the-art performance, outperforming the strongest baseline (CausalVAE) in balanced accuracy by 4.1\%, 3.8\%, and 7.0\% across the three datasets. Crucially, our approach overcomes the accuracy-interpretability trade-off. Unlike traditional models (e.g., ProtoPNet, PIP-Net) that waste representation capacity on spurious artifacts, leading to both performance drops and severely degraded prototype purity (Table~\ref{tab2}), CausalProto strictly isolates confounders. This isolation minimizes mutual information and maximizes prototype purity, simultaneously ensuring high visual transparency and superior diagnostic accuracy.

\subsubsection{Ablation study:} The ablation analysis validates each architectural component across both diagnostic accuracy and feature disentanglement. Removing the mutual information penalty (w/o MI) or merging the prototype spaces (Shared Proto) triggers an NMI surge and a sharp purity drop in Table~\ref{tab2}, subsequently causing severe accuracy declines in Table~\ref{tab1}. This confirms strict feature orthogonality fundamentally drives de-biasing. Omitting the clustering constraint (w/o Cluster) drastically reduces spurious diversity, which restricts the confounder dictionary's semantic coverage and degrades performance. Finally, while bypassing the causal intervention module (w/o do-calc) maintains high disentanglement quality, its significant diagnostic drop proves that explicitly marginalizing the spurious dictionary via do-calculus during inference remains essential to fully exploit purified representations and block residual shortcuts.

\subsubsection{Visualizing causal evidence:} Fig.~\ref{fig:exp} visualizes CausalProto’s de-confounded decision-making process. The generated heatmaps reveal a stringent focus on intrinsic pathological regions, successfully bypassing the shortcut learning of ubiquitous environmental artifacts (red boxes). Beyond accurate localization, the framework offers an intuitive trace of its reasoning by retrieving causal prototypes ($P_c$) that share robust morphological similarities with the input lesion. These highly pure pathological patches serve as direct evidence for the model's high-confidence predictions. By decoupling diagnostic signals from observational biases, CausalProto yields expert-aligned explanations that bolster clinical trust.

\section{Discussion and Conclusion}

In this work, we introduce CausalProto, demonstrating that causal inference can fundamentally resolve the persistent accuracy-interpretability trade-off in automated dermoscopy diagnosis. By combining an information bottleneck with do-calculus, our framework shifts diagnosis from fitting observational biases to interventional reasoning, effectively decoupling pathological signals from environmental artifacts. While our unsupervised confounding dictionary effectively marginalizes visual noise, its reliance on image-level features intrinsically limits the capture of complex non-visual confounders~\cite{daneshjou2022disparities}, highlighting a clear trajectory for incorporating multi-modal clinical priors in future structural causal models. Ultimately, by explicitly grounding high-confidence predictions in pure, expert-aligned visual evidence, CausalProto establishes a transparent, de-biased foundation for deploying trustworthy AI in high-stakes clinical environments.

%
%
%
\bibliographystyle{splncs04}
\bibliography{mybibliography}
%




\end{document}